\documentclass[english,prl,twocolumn]{revtex4}
\usepackage[T1]{fontenc}
\usepackage[latin1]{inputenc}
\usepackage{graphicx}

\makeatletter

\providecommand{\tabularnewline}{\\}

\makeatletter

\newcommand{\be}{\begin{eqnarray}}
\newcommand{\ee}{\end{eqnarray}}

\newcommand{\mat}{\left( \begin{array}{cc}}
\newcommand{\rix}{\end{array} \right)}

\makeatother

\usepackage{babel}
\makeatother

\begin{document}

\title{Realization of quantum walks with negligible decoherence in waveguide
lattices}

\author{Hagai B. Perets$^{1*}$ Yoav Lahini$^{1}$, Francesca Pozzi$^{2}$,
Marc Sorel$^{2}$, Roberto Morandotti$^{3}$, Yaron Silberberg$^{1}$}

\email{hagai.perets@weizmann.ac.il}

\affiliation{$^{1}$Faculty of physics, the Weizmann Institute of Science, Rehovot,
Israel}

\affiliation{$^{2}$Department of Electrical and Electronic Engineering, University
of Glasgow, Glasgow, Scotland}

\affiliation{$^{3}$Institut national de la recherche scientifique, Universit$\acute{e}$
du Qu$\acute{e}$bec, Varennes, Qu$\acute{e}$bec, Canada}

\begin{abstract}
Quantum random walks are the quantum counterpart of classical random
walks, and were recently studied in the context of quantum computation.
A quantum random walker is subject to self interference, leading to
a remarkably different behavior than that of classical random walks
such as ballistic propagation or localization due to disorder. Physical
implementations of quantum walks have only been made in very small
scale systems severely limited by decoherence. Here we show that the
propagation of photons in waveguide lattices, which have been studied
extensively in recent years, are essentially an implementation of
quantum walks. Since waveguide lattices are easily constructed at
large scales and display negligible decoherence, they can serve as
an ideal and versatile experimental playground for the study of quantum
walks and quantum algorithms. We experimentally observe quantum walks
in large systems ($\sim100$ sites) and confirm quantum walks effects
which were studied theoretically, including ballistic propagation,
disorder and boundary related effects. 
\end{abstract}
\maketitle
In classical random walks, a particle starting from an initial site
on a lattice randomly chooses a direction, and then moves to a neighboring
site accordingly. This process is repeated until some chosen final
time. This simple random walk scheme is known to be described by a
Gaussian probability distribution of the particle position, where
the average absolute distance of the particle from the origin grows
as the square root of time. First suggested by Feynman \citep{fey+65}
the term \emph{quantum} random walks was defined to describe the random
walk behavior of a quantum particle. The coherent character of the
quantum particle plays a major role in its dynamics, giving rise to
markedly different behavior of quantum walks (QWs) compared with classical
ones. For example, in periodic systems, the quantum particle propagates
much faster than its classical counterpart, and its distance from
the origin grows linearly with time (ballistic propagation) rather
then diffusively \citep{2003ConPh..44..307K}. In disordered systems,
the expansion of the quantum mechanical wave-function can be exponentially
suppressed even for infinitesimal amount of disorder, while such suppression
does not occur in classical random walks. 

In recent years QWs have been extensively studied theoretically \citep{2003ConPh..44..307K}
and have been used to devise new quantum computation algorithms \citep{2006quant.ph..9035K}.
Both discrete and continuous time QWs (DQWs;CQWs) \citep{1992JChPh..97.5148G,1993PhRvA..48.1687A,PhysRevA.58.915}
have been studied. In DQWs the quantum particle hops between lattice
sites in discrete time steps, while in CQW the probability amplitude
of the particle leaks continuously to neighboring sites. Both types
of QWs have been studied theoretically. Experimentally, many methods
have been suggested for the implementation of DQWs \citep[see][]{2003ConPh..44..307K},
but only a small scale system consisting of a few states was implemented,
using linear optical elements \citep{Do+05}. For CQWs, a few suggestions
have been made \citep{2006NJPh....8..156C,2007arXiv0705.3700M}, yet
only one experimental method have been implemented by realizing a
small scale cyclic system (4 states) using a nuclear magnetic resonance
system\citep{2003PhRvA..67d2316D}. Such systems are difficult to
scale to much larger configurations. Moreover, even at these very
small scales, errors attributed to decoherence have been observed. 

Here we suggest a very different implementation of CQWs using optical
waveguide lattices. These systems have been studied extensively in
recent years \citep{2003Natur.424..817C}, but not in the context
of QWs and quantum algorithms. We show that these systems can serve
as a unique and robust tool for the study of CQWs. For this purpose
we demonstrate three fundamental QW effects that have been theoretically
analyzed in the QW literature. These include ballistic propagation
in the largest system reported to date ($\sim100$ sites); the effects
of disorder on QWs; and QWs with reflecting boundary conditions (related
to Berry's {}``particle in a box'' and quantum carpets \citep{1996JPhA...29.6617B,2005PhRvL..95e3902I}).
Waveguide lattices can be easily realized with even larger scales
than shown here ($10^{2}-10^{4}$ sites with current fabrication technologies),
with practically no decoherence. The high level of engineering and
control of these systems enable the study of a wide range of different
parameters and initial conditions. Specifically it allows the implementation
and study of a large variety of CQWs and show experimental observations
of their unique behavior.

The CQW model was first suggested by Farhi and Gutmann \citep{PhysRevA.58.915},
where the intuition behind it comes from continuous time classical
Markov chains. In the classical random walk on a graph, a step can
be described by a matrix $M$ which transforms the probability distribution
for the particle position over the graph nodes (sites). The entries
of the matrix $M_{j,k}$ give the probability to go from site $j$
to site $k$ in one step of the walk. The idea was to carry this construction
over to the quantum case, where the {\em Hamiltonian} of the process
is used as the generator matrix. The system is evolved using $U(t)=\exp(-iHt)$.
If we start in some initial state $\left|\Psi_{in}\right\rangle $,
evolve it under $U$ for a time $T$ and measure the positions of
the resulting state, we obtain a probability distribution over the
vertices of the graph. This is described by\begin{equation}
i\frac{\partial\psi_{j}}{\partial t}=-d_{j}\gamma\psi_{j}+\gamma_{j,j+1}\psi_{j+1}+\gamma_{j,,j-1}\psi_{j-1},\label{eq:qrw}\end{equation}
 where $\psi_{j}$ is the wave function at site $j$, $d_{j}$ is
the number of sites connected to site $j$ ($d_{j}=2$ in the 1D nearest
neighbor case), and $\gamma_{i,j}(=\gamma_{j,i})$ is the probability
per unit time for the transition between neighboring \citep{Feynman,citeulike:495035}.
This mathematical formulation is effectively identical to the well
known discrete Schrödinger equation used in the tight binding (Bloch
ansatz) formalism in solid state physics \citep{citeulike:495035}.
It is used to describe the evolution of a wave-function on a periodic
potential, which is essentially the propagation of a quantum particle
on a lattice \citep{2005PhRvE..71c6128M,2007PhRvA..76a2315K}.

An immediate implication for the correspondence between QWs and these
processes is that many of the experiments in solid state physics described
by the tight-binding model could serve as implementations of QWs.
However, such experiments deal with the macro-physics of the system
and with overall observables such as conductance or transmission.
Therefore, one can not measure the specific spatial and temporal distribution
of the electrons or photons wave-functions and the micro-physics of
the system can not be directly observed. Moreover, solid state systems
contain many electrons which interact non-trivially and thus can not
be described by the evolution equation of a single particle usually
studied in QWs. Consequently, a qualitatively different experimental
approach is needed in order to study QWs. Here we report such an approach
using waveguide lattices.

\begin{figure}[t]
\begin{tabular}{c}
\includegraphics[clip,scale=0.9]{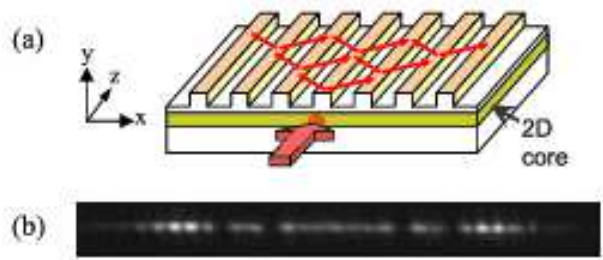}\tabularnewline
\end{tabular}

\caption{\label{f:apparatus} (a) Schematic view of the optical waveguide lattice
used in the experiments (see text). (b) Image of the output light
distribution as recorded in the infrared camera, when the light is
injected to a single lattice site at the input.}

\end{figure}

Recently, a new technique has been developed for the experimental
investigation of periodic systems using optics. The salient feature
of these experiments is that evolution of waves in time is also spread
out in space, making it much easier to observe. This is achieved by
using waveguide structures which are periodic on one dimension (x-axis;
see Fig. \ref{f:apparatus}a), but are homogeneous along the other
(z-axis). In this way the wave propagation along the z-axis is free
and corresponds to the evolution in time \citep{2003Natur.424..817C}.
Under appropriate conditions light is guided inside the waveguides
and can coherently tunnel between them. The experimental setup and
typical lattice parameters are described elsewhere\citep{1998PhRvL..81.3383E}.

Light propagating in weakly coupled, single mode waveguides, can be
described by \citep{yar89}: \begin{equation}
i\frac{n}{c}\frac{\partial A_{j}}{\partial t}=i\frac{\partial A_{j}}{\partial z}=\beta_{j}A_{j}+C_{j,j+1}A_{j+1}+C_{j,j-1}A_{j-1}\label{eq:DSE}\end{equation}

Here $A_{j}$ is the wave amplitude at site $j$, $\beta_{j}$ is
the on-site eigenvalue, $C_{i,j}$ is the coupling constant or tunneling
rate between two adjacent sites i and j (for a periodic lattice $C_{i,j}\equiv C$
is constant), and $z$ is the longitudinal space coordinate. The description
by Eq. \ref{eq:DSE} is completely analogous to the quantum description
of non-interacting electrons in a solid crystal in the tight binding
approximation, i.e. the discrete Schrödinger equation. The main differences
are that (1) the spatial modulation of the index of refraction in
the $x$ direction now plays the role of the tight binding potential,
and the $\beta_{j}$s represent the propagation-constant eigenvalues
of each waveguide in the lattice (2) the evolution at a given time
can be observed by measuring the intensity distribution at the corresponding
position in the $z$ - axis \citep{2003Natur.424..817C}, since $z=ct/n$,
where $c/n$ is the speed of light in the medium. The advantage of
this system is the possibility to control the exact initial conditions
for the light propagating inside the lattice. This is done by setting
the width, the phase and the position across the lattice of the beam
injected into the structure. In addition, this approach enables direct
observation of the resulting wave-function by measuring the distribution
of light intensity at the sample's output (fig. \ref{f:apparatus}b).
Furthermore, the temporal evolution of the wave-function can be observed
by changing the sample length, or the initial conditions (e.g. \citep{1999PhRvL..83.4756M,2005PhRvL..95e3902I}). 

\begin{figure}[t]
\begin{tabular}{c}
\includegraphics[clip,scale=0.85]{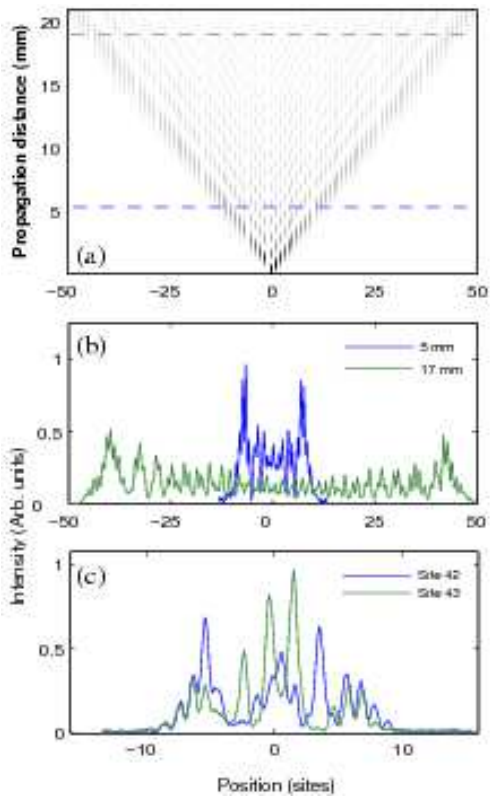}\tabularnewline
\end{tabular}

\caption{\label{f:ballistic} (a) The theoretical prediction showing the ballistic
evolution of the probability distribution of a CQW. The dashed lines
correspond to the experimental measurements in (b). (b) The observed
output pattern of light intensity after short (blue) and long (green)
propagation in a periodic lattice. This well known pattern is one
of the hallmarks of the ballistic propagation of QWs. (c) Output patterns
of light intensity resulting from injection of light into two adjacent
single waveguides (sites 42 and 43) of a disordered lattice. The different
patterns observed demonstrate the high sensitivity of the QW to the
initial conditions in this case. }

\end{figure}

One of the hallmarks of QWs on ordered lattices is their ballistic
propagation \citep{2003ConPh..44..307K}. In order to observe this
behavior, coherent light is injected into a single site in the lattice
and the output intensity is measured. In Fig. \ref{f:ballistic} we
compare the theoretical and the measured output distribution. The
signature of ballistic propagation is clearly observed both at short
and long propagation times (Fig. \ref{f:ballistic}a). Note that decoherence
effects are negligible even after relatively long evolution in time,
maintaining the detailed interference pattern predicted by theory
(Fig. \ref{f:ballistic}b). Similar results, studied in a different
context, were observed as early as in 1973 by Somekh et al. \citep{Som+73}
on small scales in structures similar to the ones described above.
 The propagating photons tunnel from the origin site to an adjacent
site, and immediately start tunneling to the next neighboring site.
Through the tunneling between sites the photons accumulate a $\pi/2$
phase, and an additional phase is accumulated continuously in each
lattice site $j$, at a rate given by $\beta_{j}$. The interference
of all these waves depends on the phase accumulated in each possible
path, and gives rise to the observed intensity distribution. This
description is practically identical to the description of the QW,
where the light intensity corresponds to the probability distribution
of the quantum particle. Since the single photon and many photon problems
are described by the same probability distribution, experiments measuring
light intensity are equivalent to performing a series of single photon
experiments, from which the probability distribution is obtained.
The propagation of more complex quantum states can be studied using
correlated or entangled photons (see for example \citep{Pol+08}).
In this case the particle characteristics of the quantum walkers can
be revealed by measuring two-photon correlation functions.

When disordered lattices are used \citep{segev,Lah+07}, a different
behavior is observed. Accumulated random phases of the random walker
lead to destructive interferences that increase with distance from
the origin. As a result, after a short ballistic propagation, the
tails of the distribution are exponentially suppressed leaving the
probability distribution exponentially localized to a small regime.
This phenomena should be distinguished from a disordered related decoherence.
Decoherence is related to temporal disorder, which induces a loss
of phase coherence and results in a transition into classical diffusion,
characterized by an expanding Gaussian probability distribution \citep{2006quant.ph..6016K,yin:022302}.
Spatial disorder such as used here leads to an exponential (Anderson)
localization \citep[e.g. ][]{Anderson,she89}, which is a coherent
interference effect . In the context of CQWs, such behavior was was
found to be important for the efficiency of quantum algorithms \citep{2007PhRvA..76a2315K,2007quant.ph..1034M,yin:022302}. 

QWs in disordered lattices are highly sensitive to the initial conditions.Fig.
\ref{f:ballistic}c shows two output patterns of light intensity resulting
from the injection of light into a single waveguide of a lattice and
similar injection to an adjacent site of the same lattice. The different
patterns observed demonstrate the high sensitivity of the QW to the
exact initial conditions. This serves as a unique signature of the
coherent nature of the QW, which is not present in the classical case.
In addition these results demonstrate the effect of disorder on QWs,
where in this case the disorder was introduced through randomizing
the tunneling rate between sites (off-diagonal disorder). The tails
of the distribution still show the ballistic component of the regular
QW. However, additional strong peaks now appear near the origin. At
later times these peaks evolve (on average) into an exponentially
localized distribution, while the ballistic side lobes are suppressed
(see \citep{Lah+07} for detailed discussion). 

\begin{figure}[t]
\begin{tabular}{c}
\includegraphics[clip,scale=0.9]{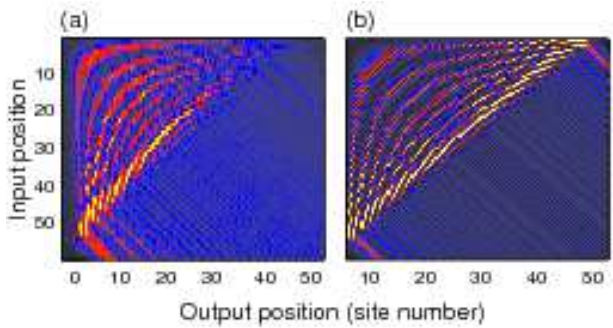}\tabularnewline
\end{tabular}

\caption{\label{f:boundry} (a) Measurements of the self interference patterns
of QWs near a reflecting boundary. Horizontal crossections show the
left half of the probability distribution of the QW, at decreasing
input site position (vertical axis), where position 0 marks the lattice
left boundary. (b) Comparison to the theoretical analysis using the
method of images \citep{2006PhRvE..73c6616M}.}

\end{figure}

Several theoretical studies have been done on QWs with boundary conditions
\citep{2004PhRvB..69s5107A,2006quant.ph..9128B}, that give rise to
complex self interference patterns. In Fig. \ref{f:boundry} we show
experimental results of a QW with one reflecting boundary condition,
compared with the theoretical analysis. A series of measurements is
shown (horizontal crossections), where in each measurement light was
injected closer to the boundary. The observed pattern results from
the self interference of the incoming and reflected photons near the
boundary, in agreement with theoretical predictions \citep{2004PhRvB..69s5107A,2006PhRvE..73c6616M}.
Although these are limited observations showing results of a short
time propagation, longer waveguide lattices could be used to study
the more complex evolution at later times. For example, such behavior
of a two boundary conditions system can be used for studying quantum
carpets containing fractal patterns \citep{1996JPhA...29.6617B,2007quant.ph..1034M}.

As an implementation of QWs, waveguide lattices carry some important
advantages over other possible schemes. First, the technologies available
for their fabrication or induction have reached a peak in recent years,
enabling full control of every lattice parameters in 1D and 2D \citep{1998PhRvL..81.3383E,2005OExpr..1310552S}
, or limited yet real time control of lattice parameters in 2D \citep{2003Natur.422..147F}.
Second, waveguide lattices have excellent structural stability, thus
in practice decoherence due to noise is negligible. The optical wavelength
in our experiments (using AlGaAs wafers) is around 1.5$\mu$m, the
standard communication wavelength, and losses at these wavelengths
are extremely small. This is highly important for quantum computational
tasks where coherency is essential. Third, effects arising from the
interactions between different random walkers in other possible implementations
are eliminated here, due to the bosonic, non interacting nature of
photons. 

In recent years several quantum algorithms based on QWs have been
suggested\citep{2004quant.ph..3120A}. For realistic use of such algorithms
one requires exponentially large systems. We note that as long as
entanglement is not introduced, our system is limited to large but
not exponentially large scale functionality. The lack of entanglement
limits the number of the states of the system, which scales linearly
with the number of waveguides. Our system, even without entanglement,
can potentially implement QW algorithms, since quantum entanglement
is not required for the algorithm implementation or its improved efficiency.
Its only role in this case is to allow for a larger number of states
(see, for example the discussion in \citep{2002PhRvL..88m7901B}).
Some of the suggested QW algorithms have been shown to provide polynomial
or even exponential speed up \citep{2002quant.ph..9131C,2004PhRvA..70b2314C}.
Unfortunately, in all of the algorithms suggested so far the speed
up of quantum over classical algorithms is achieved only when applied
to high dimensional systems. Nevertheless, our system can still be
used to implement and study these algorithms in lower dimensions. 

In summary, we have demonstrated the strong correspondence between
QWs and light propagation in waveguide lattices. This correspondence
can be used to extend and interchange ideas and knowledge acquired
in both fields (e.g. non-linear behavior \citep{2003Natur.424..817C}
in CQWs or entanglement effects \citep{2005NJPh....7..156C,2006PhRvA..73d2302A}
in waveguide lattices). The high level of control, the accuracy, and
the low decoherence rates achieved in waveguide lattices experiments
provide a powerful tools for the study of QWs, and may prove useful
in the implementation of QWs-based algorithms.

This work was supported by the German-Israeli Project Cooperation
(DIP), NSERC and CIPI (Canada), and EPRSC (UK). Y.L. is supported
by the Adams Fellowship Program of the Israeli Academy of Sciences
and Humanities. H.P. and Y.L. thank the members of the WIS PIRATE
club in which this study was initiated.

\bibliographystyle{apsrev} \bibliographystyle{apsrev}
\bibliography{quant.v22}

\end{document}